%% file: spie2004-pasch-arx.tex
\documentclass[]{spie}  %
\usepackage[]{graphicx}
\usepackage{url}
\newcommand{\grad}{\ensuremath{^{\circ}}} 
\newcommand{\unit}[1]{\,\ensuremath{\mathrm{#1}}}

\title{LLiST - a new star tracker camera for tip-tilt correction at IOTA
\footnote{\input{spie2004-copyright}}%
}

\author{%
P.\ A.\ Schuller\supit{a}, M.\ G.\ Lacasse\supit{b}, 
D.\ Lydon\supit{c}, W.\ H.\ McGonagle\supit{d}, 
E.\ Pedretti\supit{e}, R.\ K.\ Reich\supit{d}, \\
F.\ P.\ Schloerb\supit{c}, and W.\ A.\ Traub\supit{a}
\skiplinehalf
\supit{a}
Harvard-Smithsonian Center for Astrophysics, 
60 Garden St., Cambridge, MA 02138, USA; \\
\supit{b}
Fred Lawrence Whipple Observatory, 
P.O. Box 97, Amado, AZ 85645, USA; \\
\supit{c}
Department of Astronomy, University of Massachusetts, 
Amherst, MA 01003, USA; \\
\supit{d}
Advanced Imaging Technology Group, MIT Lincoln Laboratory, 
Lexington, MA 02420, USA; \\
\supit{e}
University of Michigan, 
941 Dennison Building, 500 Church St., Ann Arbor, MI 48109, USA
}   %

\authorinfo{%
Send offprint requests to: P.\ Schuller, 
e-mail: \texttt{pschuller@cfa.harvard.edu} 
}   %
\begin{document} 
  
  \maketitle 

\begin{abstract}
The tip-tilt correction system at the Infrared Optical Telescope Array (IOTA) 
has been upgraded with a new star tracker camera. 
The camera features a backside-illuminated CCD chip 
offering doubled overall quantum efficiency 
and a four times higher system gain compared to the previous system. 
Tests carried out to characterize the new system 
showed a higher system gain
with a lower read-out noise electron level. 
Shorter read-out cycle times now allow to compensate tip-tilt fluctuations 
so that their error imposed on visibility measurements becomes comparable to, 
and even smaller than, that of higher-order aberrations. 
\end{abstract}


\section{INTRODUCTION}
\label{sect.intro}  %
The Infrared Optical Telescope Array (IOTA) 
is located at the Smithsonian Institution's Whipple Observatory 
on Mount Hopkins, Arizona\cite{cit.traub2004SPIE}. 
It operates with three telescopes 
and offers multiple baselines in the range 5 to $38\unit{m}$. 
Typical site conditions, observed at the MMT Observatory\cite{cit.mmto-seeing} 
with a central wavelength $\lambda_c = 0.64\unit{\mu m}$, 
show a median seeing $\alpha=0.77''$ (FWHM) 
and a median (ground) wind speed 
$v_\mathrm{w}=12\unit{mph}=5.3\unit{m}\unit{s}^{-1}$. 
Both these numbers can become at least 2 times larger. 
They also allow the derivation\cite{aoarticle,AO2} of 
the Fried parameter $r_0$ and the coherence time $\tau_0$ 
which can be interpreted as scale length and time 
over which the phase fluctuations of a light wavefront coming from a star 
and entering the telescope aperture may be considered as sufficiently small. 
In a crude approximation, 
$r_0=\lambda / \alpha$ 
and 
$\tau_0=0.5\,r_0 / v_\mathrm{w}$ (assuming dominance of ground wind speed), 
which yields $r_0=17\unit{cm}$ and $\tau_0=16\unit{ms}$. 
In the H band, in which the IONIC fiber beam combiner at IOTA operates\cite{cit.berger2003,cit.ragland2004SPIE}, 
and using the fact that $r_0\propto \lambda^{6/5}$, 
$\lambda_c^\mathrm{H} = 1.65\unit{\mu m}$ so that
$r_0^\mathrm{H} = 53\unit{cm}$ and $\tau_0^\mathrm{H} = 50\unit{ms}$. 
The telescopes at IOTA have an entrance pupil $D=45\unit{cm}$ 
which is of the same order of magnitude as $r_0^\mathrm{H}$. 
Therefore, tip-tilt correction of the incoming beam 
compensates for atmospherically induced aberrations 
already to a high degree\cite{cit.shaklan1988,cit.brummelaar1995} 
where remaining visibility errors are dominated by higher-order aberrations. 

The overall procedure for tip-tilt correction that is applied at IOTA 
for each of the three telescopes is as follows. 
The visible parts of the beams are focused onto a CCD, 
i.\ e., one CCD frame shows three individual star images. 
The positions of the centroids are retrieved 
and compared to the nominal positions. 
Then the tertiary mirrors, which are mounted on 2-axis piezo driven stages, 
are adjusted accordingly. 
The whole process aims at stabilizing the 
infrared star image on the fiber input of the interferometer. 
Movements due to residual tip-tilt errors are reflected 
in varying flux recorded in the interferometer outputs. 

As the aforementioned and more detailed considerations\cite{cit.chara_tp,cit.tango1980} indicate, 
the goal is therefore to apply tip-tilt correction 
with an overall cycle time $t_\mathrm{tt} \ll \tau_0$ 
and as short as $t_\mathrm{tt}=1\unit{ms}$. 
An important unit in the tip-tilt correction system is the star tracker CCD camera. 
Its sensitivity, read-out frame rate, and read-out noise 
majorly determine if this goal can be achieved. 
Previously, the camera in use at IOTA 
could reach frame rates of $200\unit{Hz}$, 
which is a factor of 5 slower than the envisaged goal, 
and it would allow tracking stars with an apparent magnitude of up to $I=12\unit{mag}$, 
depending on integration time and seeing conditions. 
Recently, IOTA acquired a new CCD camera 
in order to reach full tip-tilt correction. 
In the following sections, we present first test results of this new camera.

\section{CAMERA PROPERTIES} 
\label{sect.cam_props}
\begin{figure}
        \centering
        \includegraphics[width=0.32\textwidth]{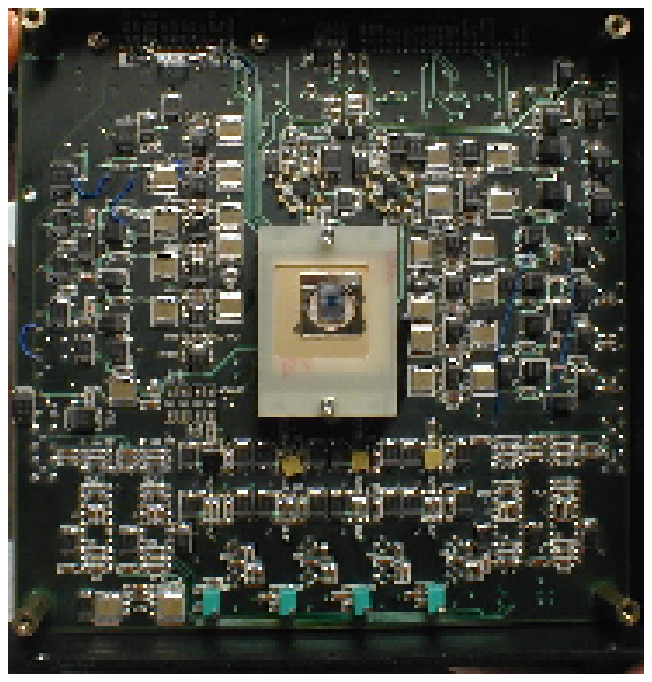} %
        \\
        \includegraphics[height=0.8\textwidth,angle=-90]{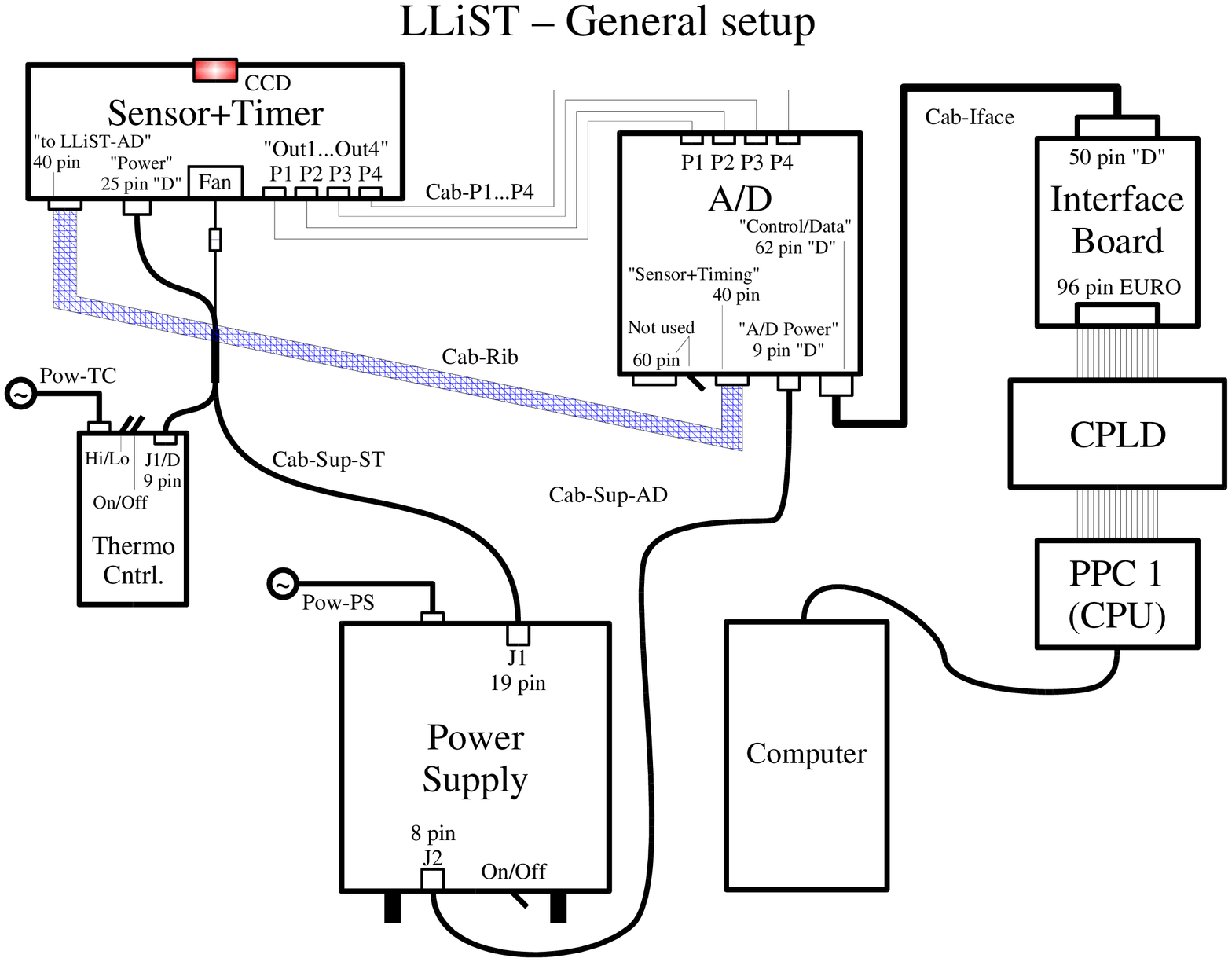} %
        \caption{%
        \label{fig.setup}
        LLiST at IOTA. 
        \emph{Top:} Electronics board holding the CCD. The board is roughly $185\unit{mm}$ on a side. 
        \emph{Bottom:} Overall setup of LLiST in the IOTA system. 
        }
        \end{figure}

The new star tracker camera for IOTA was acquired 
from MIT Lincoln Laboratory (LL) - hence the acronym LLiST. 
Its design is based on an imager 
already developed for adaptive optics applications\cite{cit.reich1993,cit.burke2000}. 
Figure~\ref{fig.setup} shows the overall setup of LLiST 
as it is implemented at IOTA. 
The CCD is kept in a sealed housing, cooled by a thermo-electric cooler (TEC) 
and mounted on the so-called sensor board. 
The timing board, which is mounted together with the sensor board in one case, 
controls the clocking of the analog and digital data 
and provides synchronization signals for the IOTA control system\cite{cit.schloerb2004SPIE}. 
A   programmable logic device (CPLD) %
controls the camera\cite{cit.pedretti2003}, 
receives data from it, and transfers data for further processing.

\subsection{Chip characteristics} 
\label{sect.chip}
\begin{figure}
        \centering
        \includegraphics[width=0.75\textwidth]{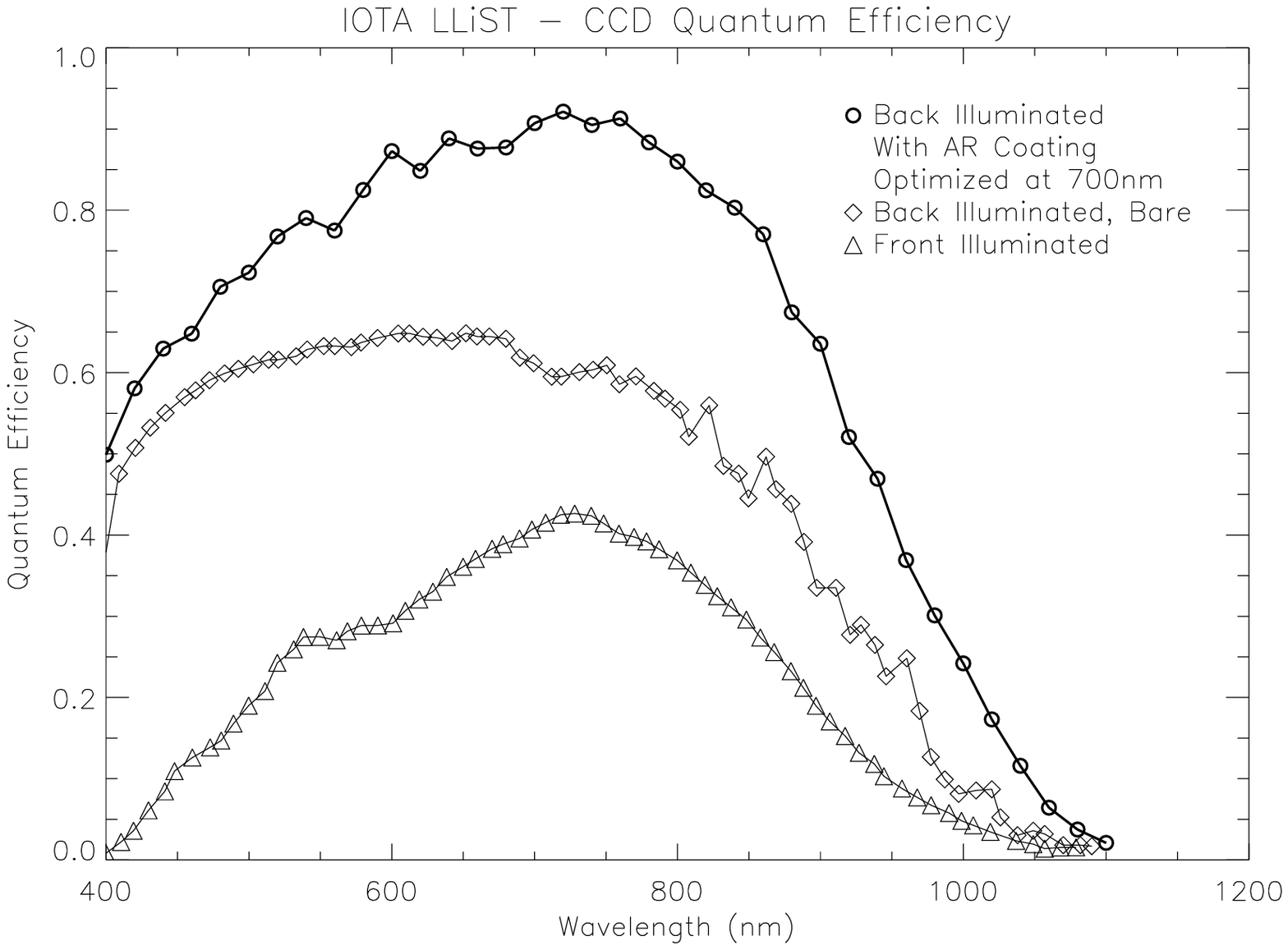} %
        \caption{%
        \label{fig.qe}
        Spectral response of the LLiST CCD\cite{cit.mit-ll} 
        (circled data points). 
        The curve reaches a maximum of 90\% 
        close to $\lambda_c=700\unit{nm}$. 
        }
        \end{figure}

The central part of LLiST is a $64\times64$-pixel backside-illuminated CCD. 
Figure~\ref{fig.qe} shows the spectral response of the chip 
showing a maximum of 90\% near $\lambda_c=0.7\unit{\mu m}$. 
A pixel is $21\unit{\mu m}$ on a side, 
with a full well capacity of about 175,000 electrons. 
The fill factor of the chip is close to $100\%$. 

At IOTA, the telescope beams are first compressed by a factor of 10 
to a diameter $d=4.5\unit{cm}$ 
and then focused on the ST camera 
by means of lenses with focal length $f_\mathrm{ST}=60\unit{cm}$. 
This yields a pixel scale on the sky of $s^\mathrm{sky}=0.72''\unit{px}^{-1}$. 
The effective full field of view (FOV) on the sky is $\phi_\mathrm{ST}^\mathrm{sky}\approx 25''$ per telescope, 
constrained by the optical train of the instrument. 
Assuming the diffraction limited case (Airy pattern), 
the central lobe of the star image on the CCD has a diameter 
$L = 2.44\cdot f_\mathrm{ST} \lambda_\mathrm{c}/d = 23\unit{\mu m}$, 
a little over one pixel.

\subsection{Read-out process} 
\label{sect.readout}
The area of the CCD exposed to light is sub-divided in four ports. 
Each port has its own analog and digital amplifier chain. 
An exposed frame is first transfered to a storage area protected from light 
and is then read out sequentially row by row into serial registers 
which are transfered to separate amplifier chains. 
In addition to receiving and transfering charge from the exposed frame, 
a serial register contains four physical pixels 
which are also shielded from light exposure. 
For this reason, 
and because of their short idle time 
($0.91\unit{\mu s}$; see clocking rate below) during read-out, 
these pixels can be used in principle as reference for the chip bias. 
When digitizing data on the analog-to-digital (AD) board, 
the order of pixels is re-arranged 
so that the reference pixels of all four ports come out last in a row. 

\begin{figure}
        \centering
        \fbox{\includegraphics[width=0.48\textwidth]{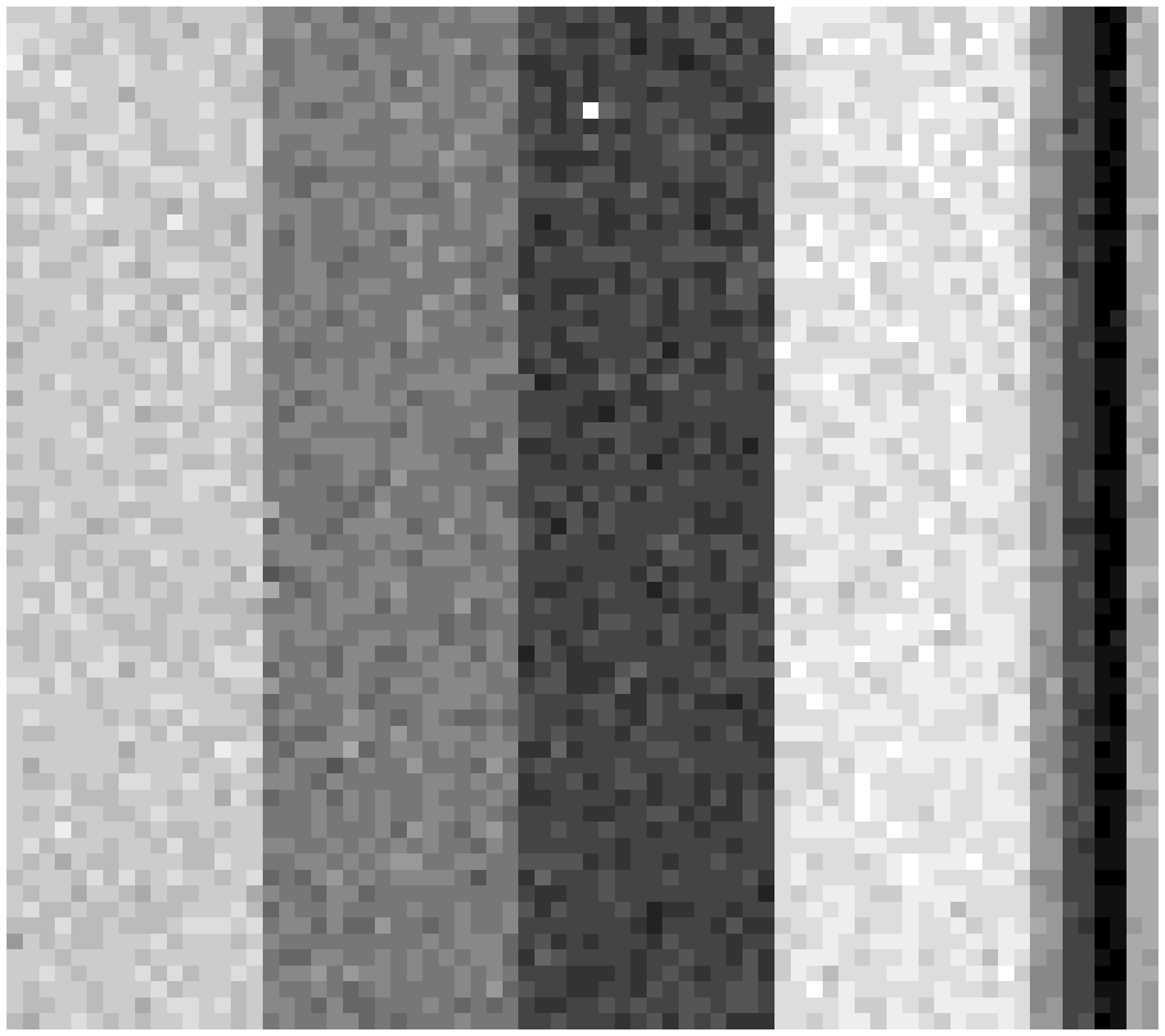}} %
        \fbox{\includegraphics[width=0.48\textwidth]{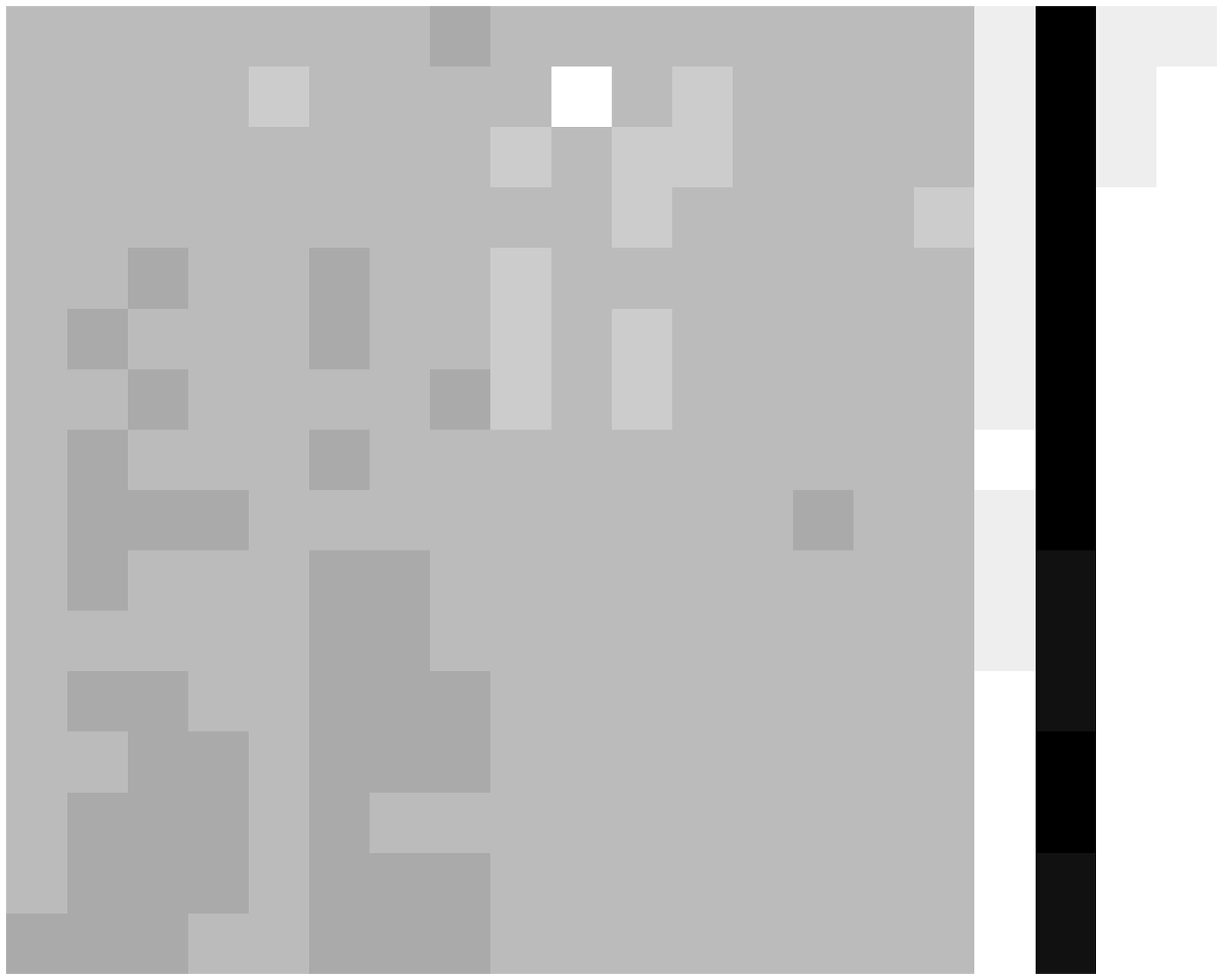}} %
        \caption{%
        \label{fig.modes}
        Sample dark frames of read-out modes M1 (left) and M2 (right). 
        In each row, the four data ports are followed by their respective reference pixels. 
        The third port from the left shows a bright pixel close to the top edge of the frame. 
        The bias to the ports was balanced in M2. Using the same settings, the bias is not balanced in M1. 
        }
        \end{figure}

The camera offers a number of different read-out modes 
of which to date mainly two have been in use. 
In Mode 1 (M1) the full frame is read out unbinned, 
whereas Mode 2 (M2) bins $4\times4$ pixels during the read-out process. 
Clocking rates of the serial register 
are $1.1\unit{MHz}$ for M1 and $0.25\unit{MHz}$ for M2. 
Maximum possible read-out frame rates on the camera side are 
$\sim 800$ frames per second (fps) in M1 and $\sim 2200\unit{fps}$ in M2, 
whereas observed cycle times of the control software 
with open star tracker control loop were 
$4.5\unit{ms}$ and $1.0\unit{ms}$, respectively. 
The limiting factor here is the data handling in the control software, 
a matter currently being worked on. 
Figure~\ref{fig.modes} shows sample frames of M1 and M2, respectively. 
Other modes read out only a window of the full binned mode, 
i.~e., a certain number of rows.

\subsection{Amplifier chain} 
\label{sect.ampl}
On the sensor board, collected charge is converted by some capacitance 
into a voltage fed to a pre-amplifier. 
This step is characterized by the responsivity $R$ of the board electronics 
($[R]$={V/e-}). 
It is not known a priori, 
but can be determined (see Sect.~\ref{sect.ron-sysgain}). 
The signal is pre-amplified with an electronic gain $G_\mathrm{e}=6.8$, 
whereby the dynamic output range of the pre-amplifier in LLiST is $2.5\unit{V}$. 
A 14-bit AD converter transfers this analog signal 
into analog-to-digital units (ADU), 
having a conversion factor $K$ 
with $K^{-1}=2.5\,\mathrm{V}/2^{14}\,\mathrm{ADU}=152.6\,\mathrm{\mu V/ADU}$. 
The overall behavior of the entire system is characterized 
by the system gain $g_\mathrm{s}$ ($[g_\mathrm{s}]=$\,{e-/ADU}) with 
        \begin{equation}
        \label{eqn.sysgain}
        g_\mathrm{s} = (RG_\mathrm{e}K)^{-1} \quad . 
        \end{equation}

\subsection{Read-out noise and system gain} 
\label{sect.ron-sysgain}
\subsubsection{Mode 1} 
\label{sect.ron-sysgain-m1}
To determine the read-out noise (RON) standard deviation $\sigma_\mathrm{ron}$ 
and the system gain $g_\mathrm{s}$ in Mode~1, 
we applied a variant of the photon-transfer technique 
as outlined in Ref.~\citenum{cit.janesick1987}. 
It uses the relation between 
the overall standard deviation $\sigma_\mathrm{tot}$ of the number of electrons collected in a pixel, 
the standard deviation $\sigma_{h\nu}$ due to photo-electron and dark current noise, 
and $\sigma_\mathrm{ron}$ in (e- RMS): 
        \begin{equation}
        \label{eqn.noise_e}
  \sigma_\mathrm{tot}^2 
      =   \sigma_{h\nu}^2 + \sigma_\mathrm{ron}^2  \quad .%
        \end{equation}
This relation is converted by the camera system into ADUs 
through its system gain $g_\mathrm{s}$: 
        \begin{eqnarray}
        \label{eqn.noise_adu} 
  (\sigma_\mathrm{tot}^\mathrm{ADU})^2  & = &
     g_\mathrm{s}^{-2} \, \sigma_{h\nu}^2 + 
     (g_\mathrm{s}^{-1} \, \sigma_\mathrm{ron})^2 \\
    & = &
     g_\mathrm{s}^{-1}\cdot [g_\mathrm{s}^{-1} \, S_{h\nu}] + 
     (g_\mathrm{s}^{-1} \, \sigma_\mathrm{ron})^2 \nonumber \\
        \label{eqn.noise_adu_sig} 
    & = &
     g_\mathrm{s}^{-1}\cdot S_{h\nu}^\mathrm{ADU} + 
     (g_\mathrm{s}^{-1} \, \sigma_\mathrm{ron})^2 \qquad .
        \end{eqnarray}
In the second to last step we inserted the signal value $S_{h\nu}$ for its variance $\sigma_{h\nu}^2$, 
assuming Poisson statistics for thermally generated and photo-electrons. 
Equation~\ref{eqn.noise_adu_sig} describes a linear relation 
between the measured signal $S_{h\nu}^\mathrm{ADU}$ 
and the total measured variance $(\sigma_\mathrm{tot}^\mathrm{ADU})^2$, 
whereby the slope represents the inverse of the system gain $g_\mathrm{s}$ 
and the intercept the squared product of $g_\mathrm{s}^{-1}$ 
and the read-out noise $\sigma_\mathrm{ron}$, 
which can also be interpreted as remaining noise at zero light level. 

Assuming that all pixels on the CCD basically behave the same way 
(i.e., common system gain $g_\mathrm{s}$) 
and follow the same statistics, 
one can expose the CCD to a non-uniform light distribution 
and take a number ($> 100$) of frames 
sufficiently high to derive statistical values from them. 
For each pixel, the signal and noise values will create a data point. 
The set of data points will scatter 
around the linear relation (\ref{eqn.noise_adu_sig}) 
so that linear regression can be applied. 

\begin{figure}
        \centering
        \fbox{\includegraphics[width=0.4\textwidth]{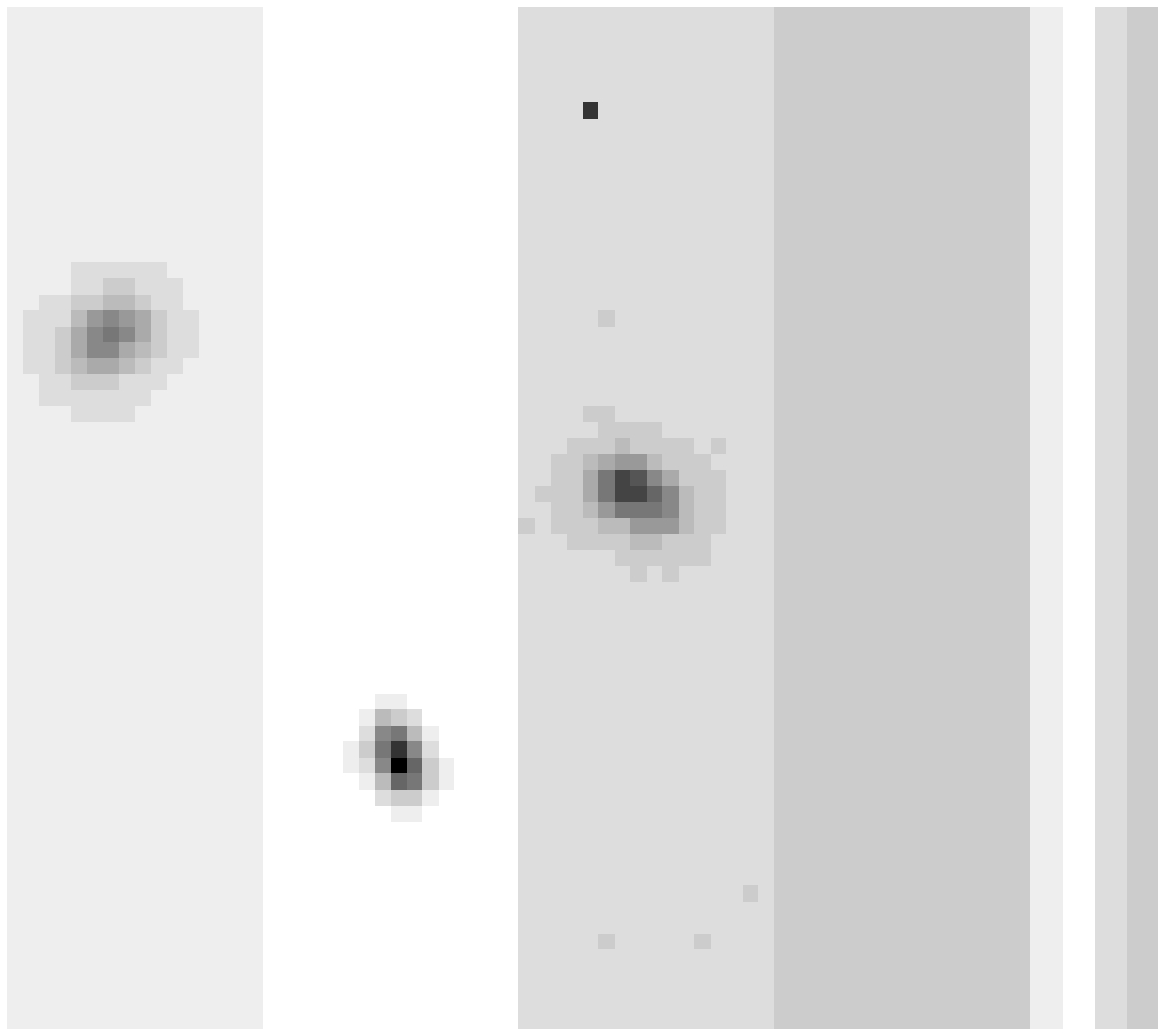}} %
        \\[1ex]
        \includegraphics[width=0.49\textwidth]{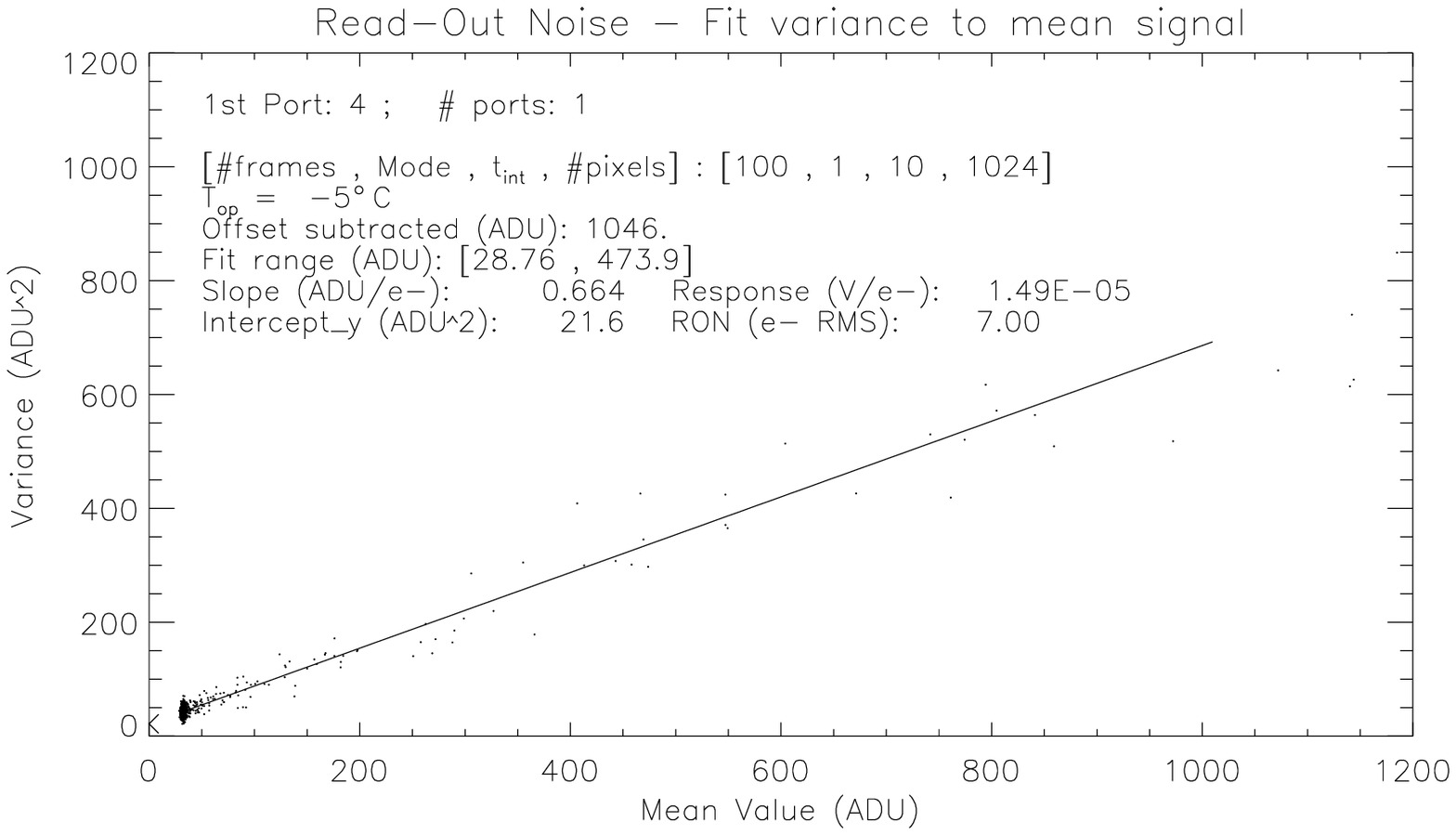}
        \includegraphics[width=0.49\textwidth]{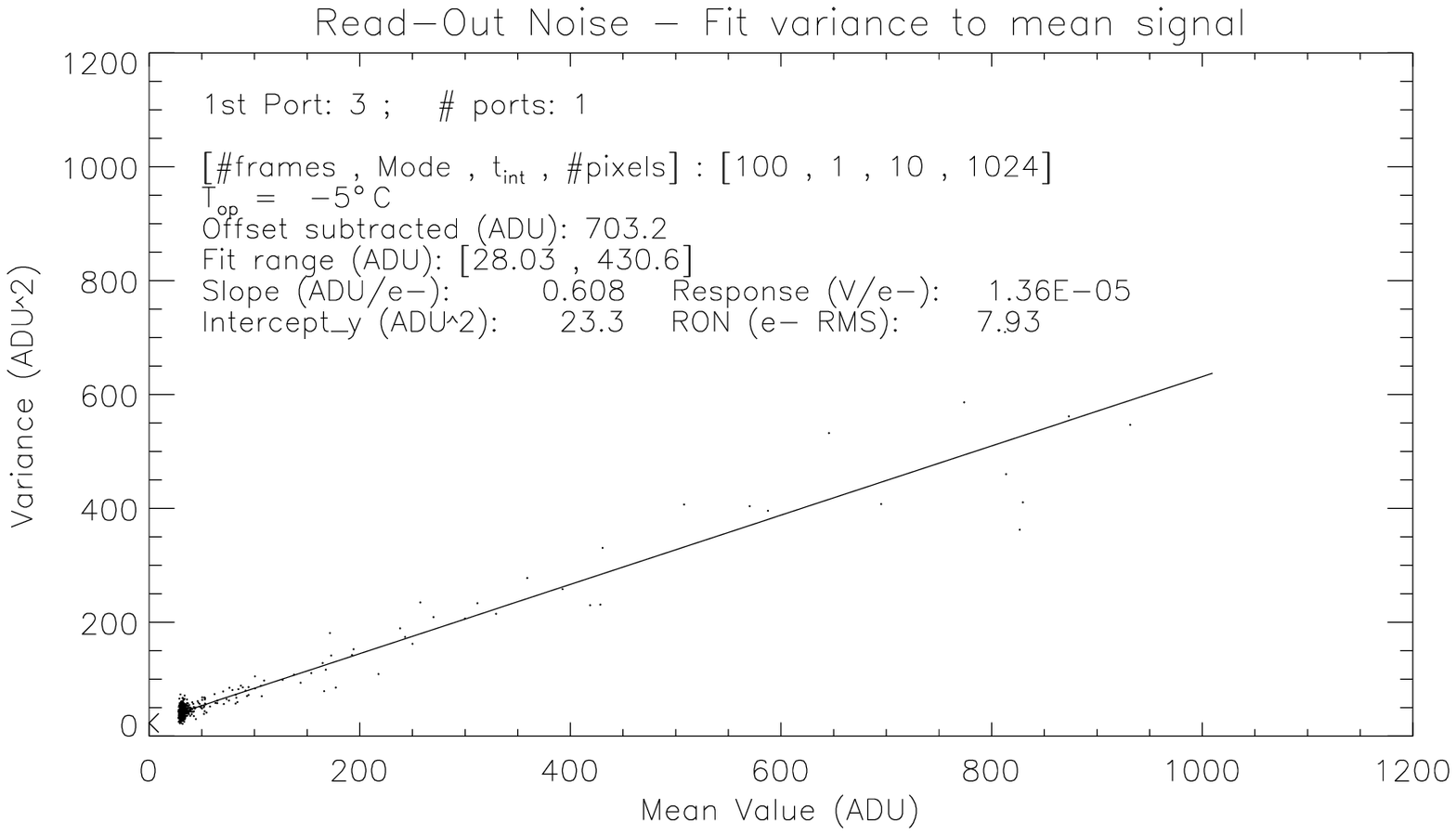}
        \\
        \includegraphics[width=0.49\textwidth]{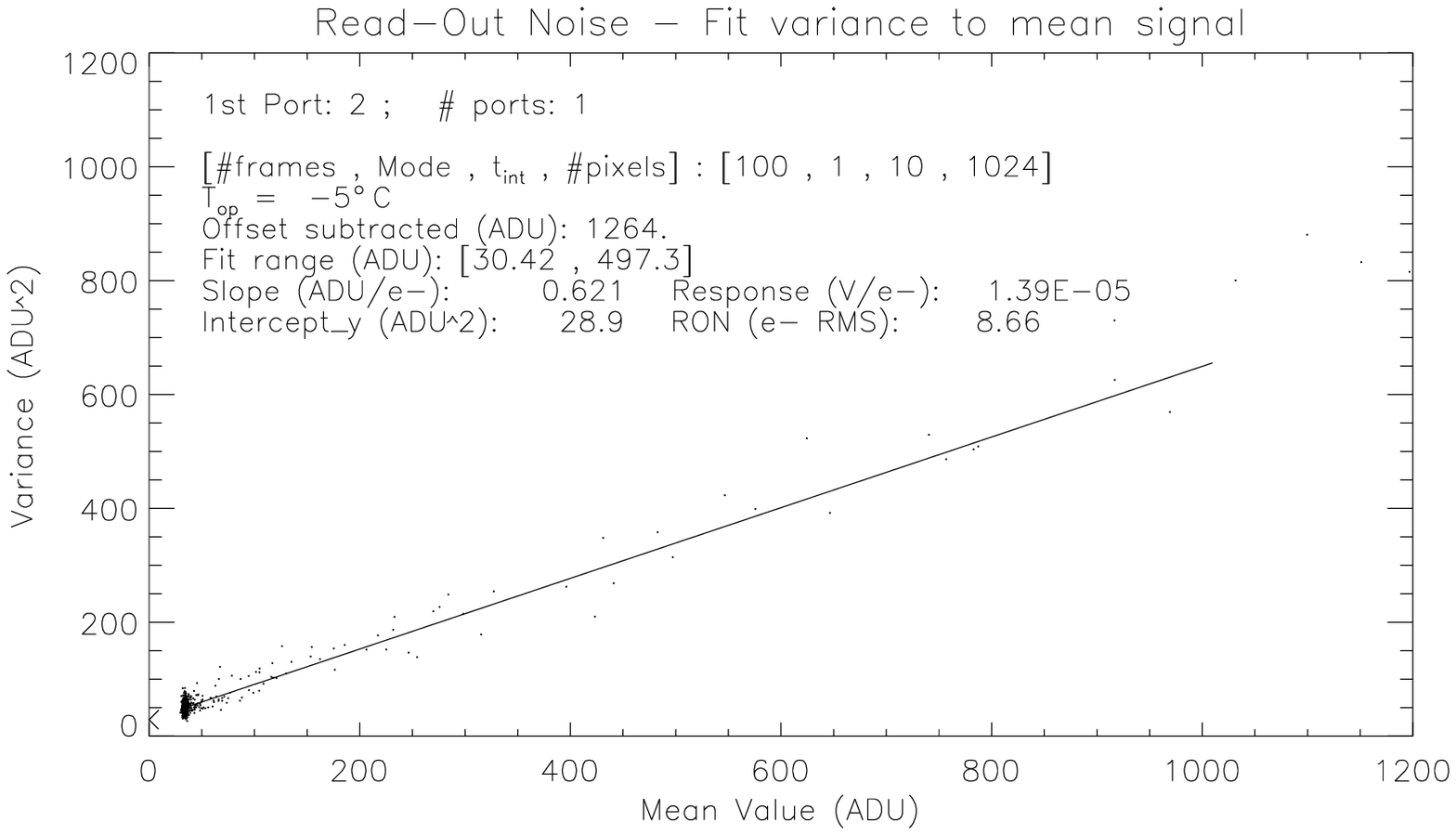}
        \includegraphics[width=0.49\textwidth]{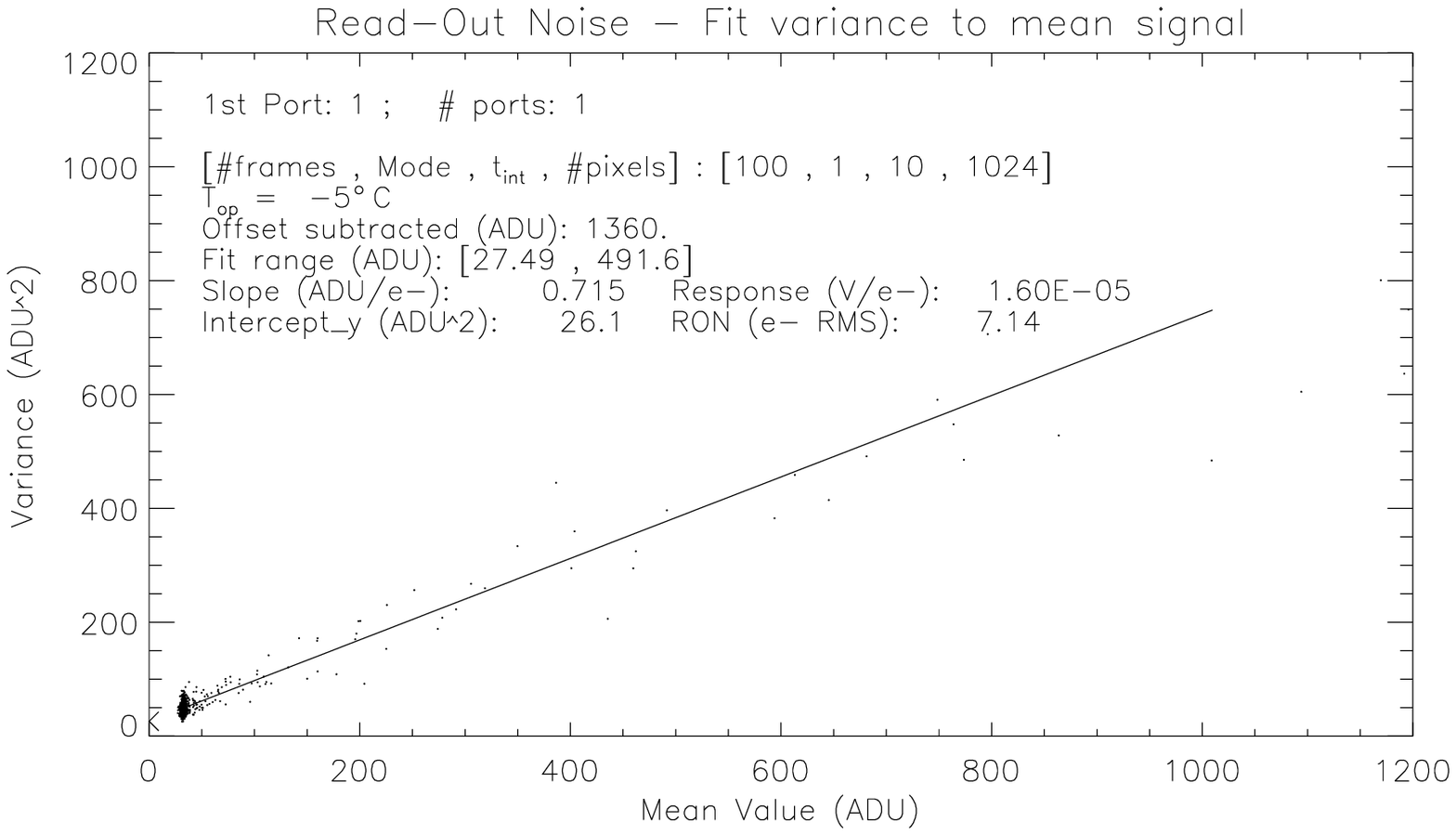}
        \caption{%
        \label{fig.phottrf}
        Application of photon-transfer technique in Mode~1 
                (see Sect.~\ref{sect.ron-sysgain-m1}). 
        \emph{Top}: A battery-powered LED back-illuminated the IONIC fibers. 
        The occuring beams were directed towards LLiST 
        and slightly defocused to emphasize non-uniform light distribution. 
        \emph{Bottom}: Results for individual ports. 
        See inserts for operational details. 
        For each port, the subtracted bias (offset) was averaged 
        over the corresponding frame reference pixels. 
        The slope of the line 
        equals the inverse of the system gain $g_\mathrm{s}^{-1}$, 
        whereas the intercept is the squared product of $g_\mathrm{s}^{-1}$ 
        and the read-out noise $\sigma_\mathrm{ron}$. 
        }
        \end{figure}

Figure~\ref{fig.phottrf} shows some results of this procedure. 
For the necessary subtraction of the camera bias, 
the reference pixels were averaged to one number %
which was subtracted off its corresponding port. 
Dark current was not subtracted, but treated as part of the signal. 
The linear fit was done at comparatively low signal levels 
with a higher data point population and where read-out noise becomes important. 

\subsubsection{Mode 2} 
\label{sect.ron-sysgain-m2}
For Mode~2, we had to apply a method 
slightly extended from the one used for Mode~1. 
Due to camera-internal reasons, 
the reference pixels in Mode~2 are not good indicators of the bias, 
they carry erroneous signal. 
Moreover, due to the $4\times4$ binning, 
the number of pixels is strongly reduced 
and therefore application of the photon-transfer technique 
would %
show scarce population along the regression line. 
Furthermore, it was not clear from the beginning 
how low the CCD had to be cooled 
so that dark current noise would not affect operation 
with intended integration times $t_\mathrm{int}$ up to 100\unit{ms}. 

For these reasons we rather took a series of dark frames with increasing $t_\mathrm{int}$ 
and treated the thermally induced dark current $I_\mathrm{d}$ as signal. 
This signal is plotted versus $t_\mathrm{int}$ 
and a linear regression is performed. 
The intercept of the ordinate yields the bias $B$ 
(``signal at zero integration time'') whereas the slope in this plot yields the dark current: 
        \begin{equation}
        \label{eqn.darkcurr}
        S_{h\nu} = I_\mathrm{d} \cdot t_\mathrm{int} + B  \quad .%
        \end{equation}
The bias was subtracted off all frames. 
These frames were then used to perform a reduction 
similar to the one described in the previous section for Mode~1. 
For Mode~2, pixel values and signal variances were averaged over a port 
to create one data point for a given integration time. 

\begin{figure}
    \centering
        \includegraphics[width=0.49\textwidth]{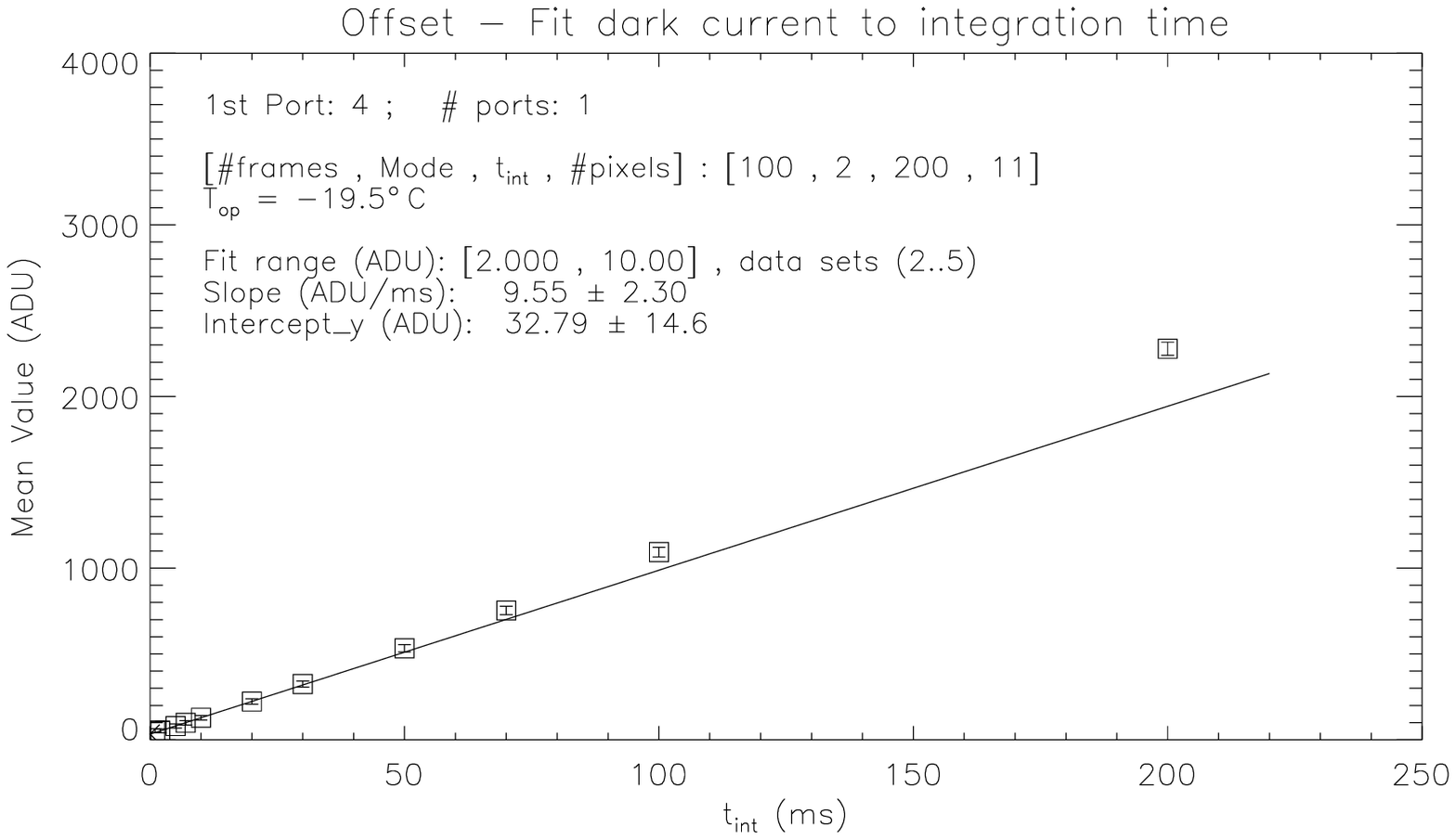}
        \includegraphics[width=0.49\textwidth]{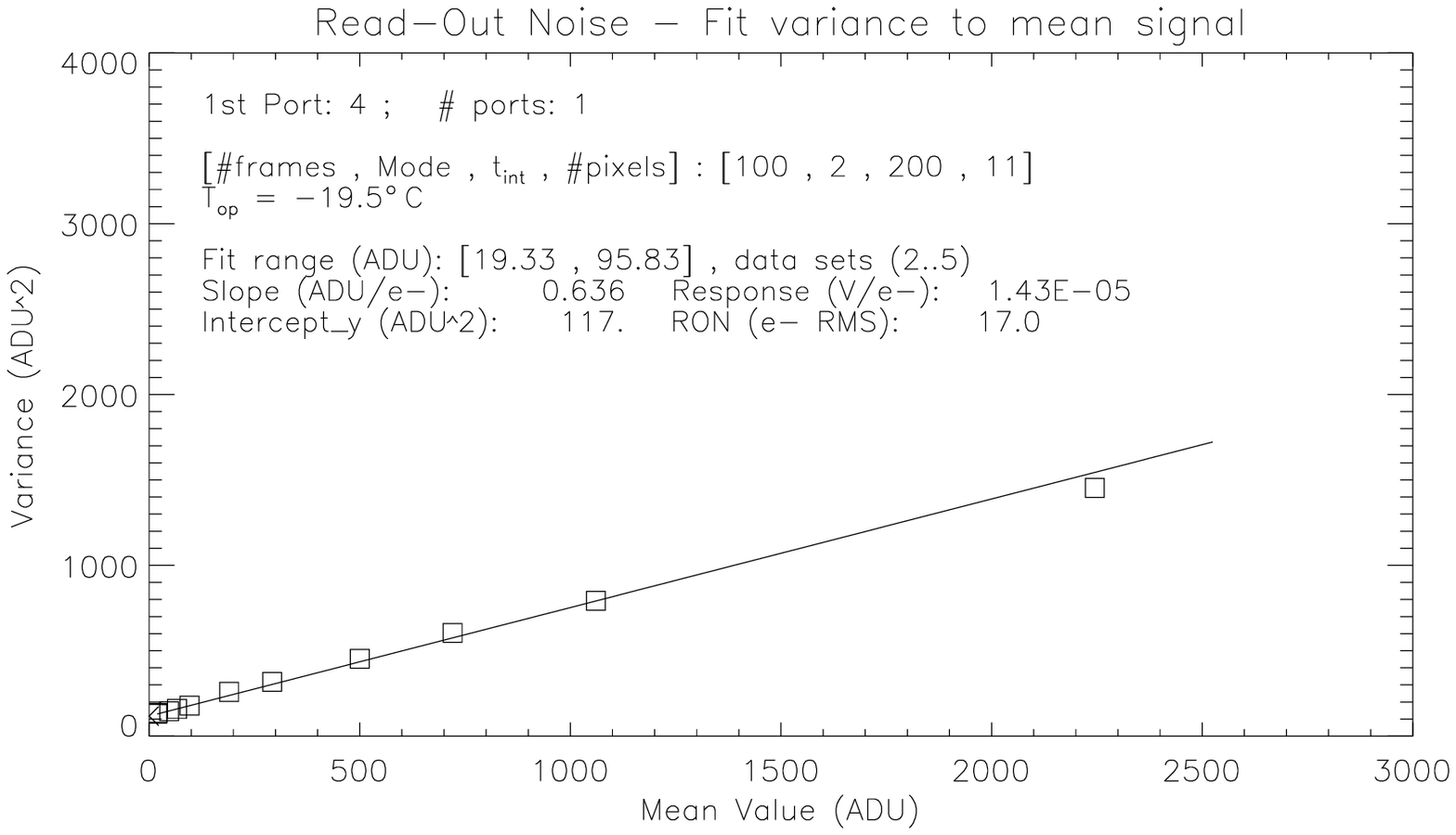}
        \\
        \includegraphics[width=0.49\textwidth]{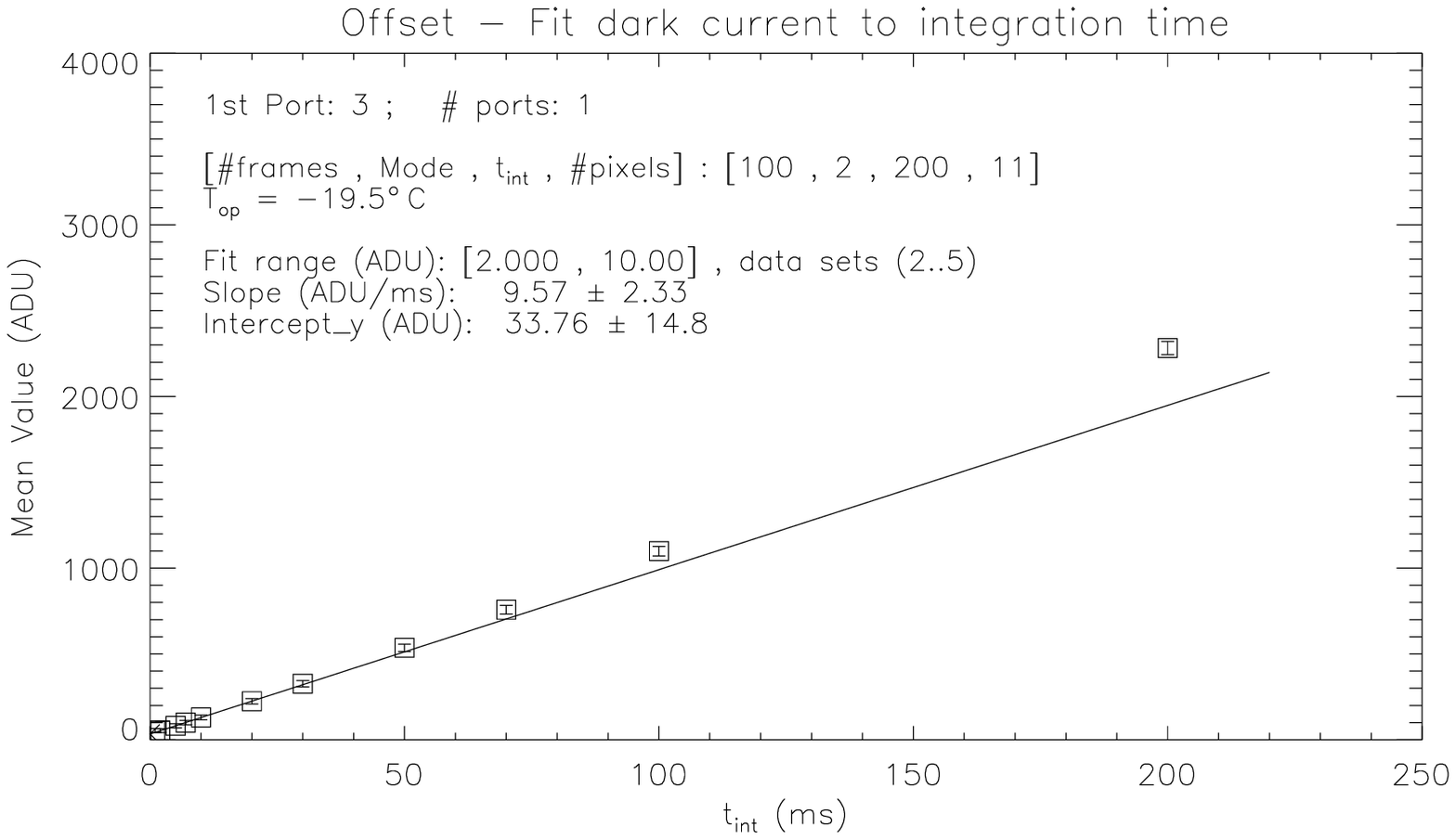}
        \includegraphics[width=0.49\textwidth]{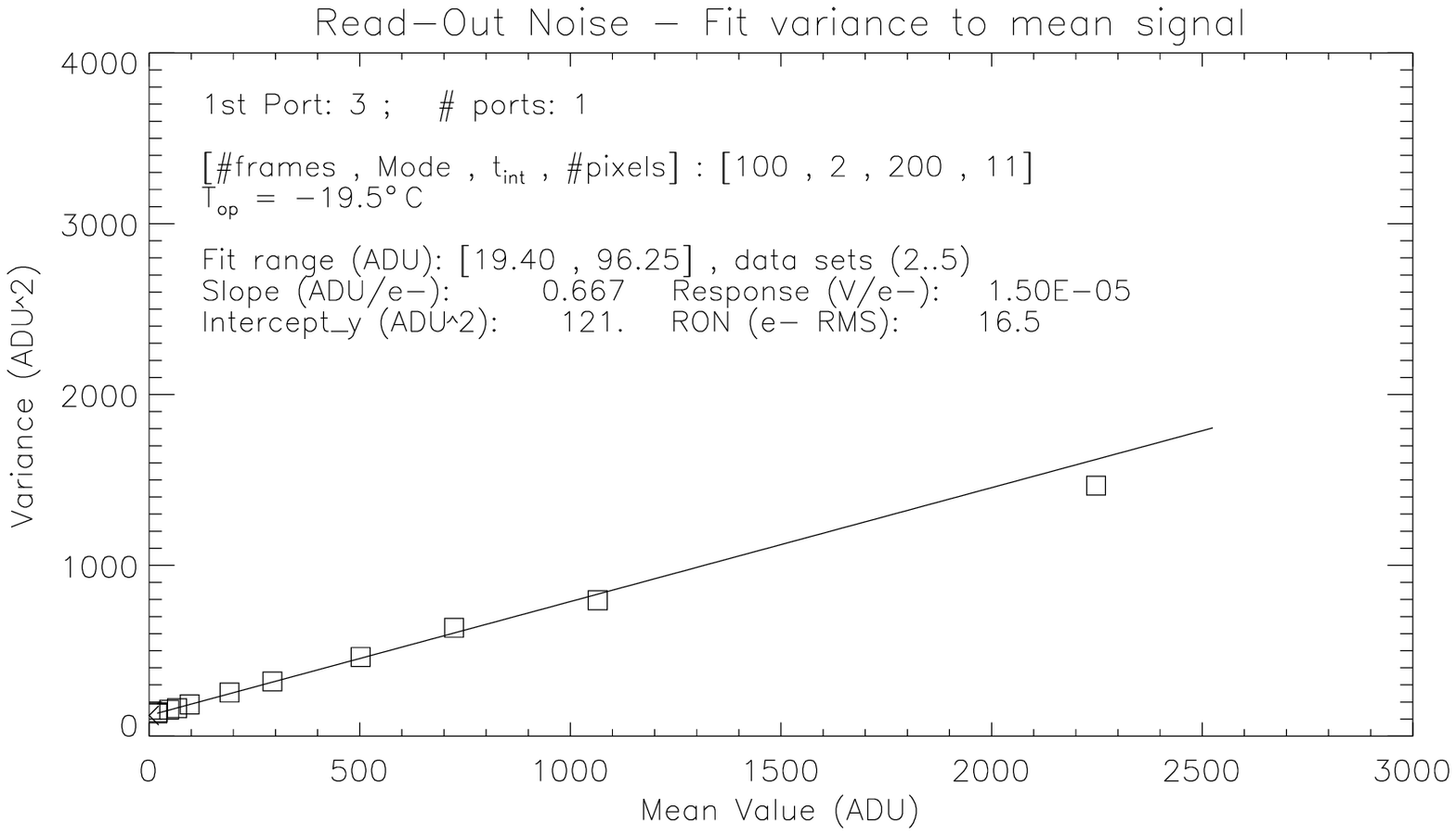}
        \\
        \includegraphics[width=0.49\textwidth]{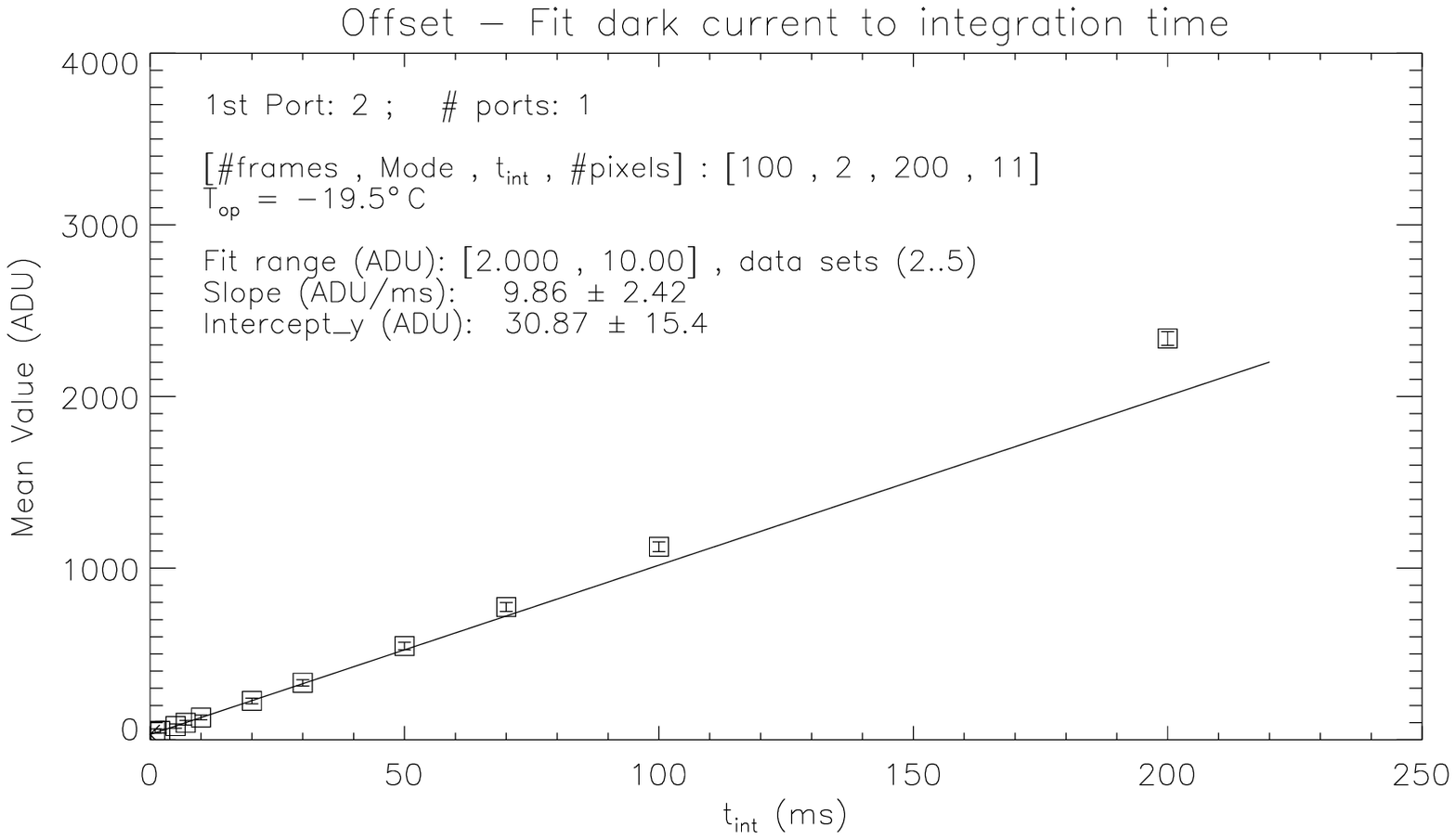}
        \includegraphics[width=0.49\textwidth]{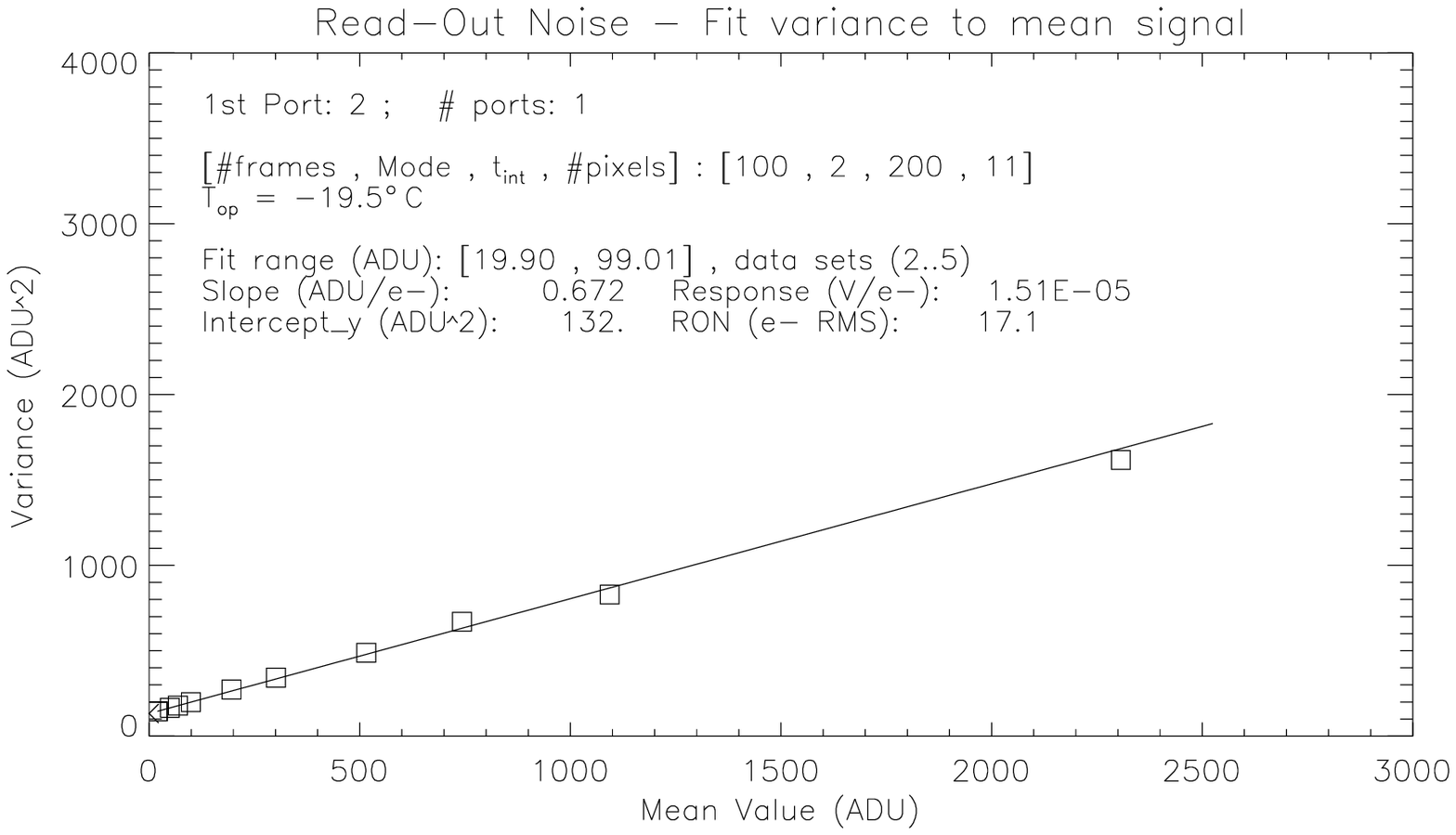}
        \\
        \includegraphics[width=0.49\textwidth]{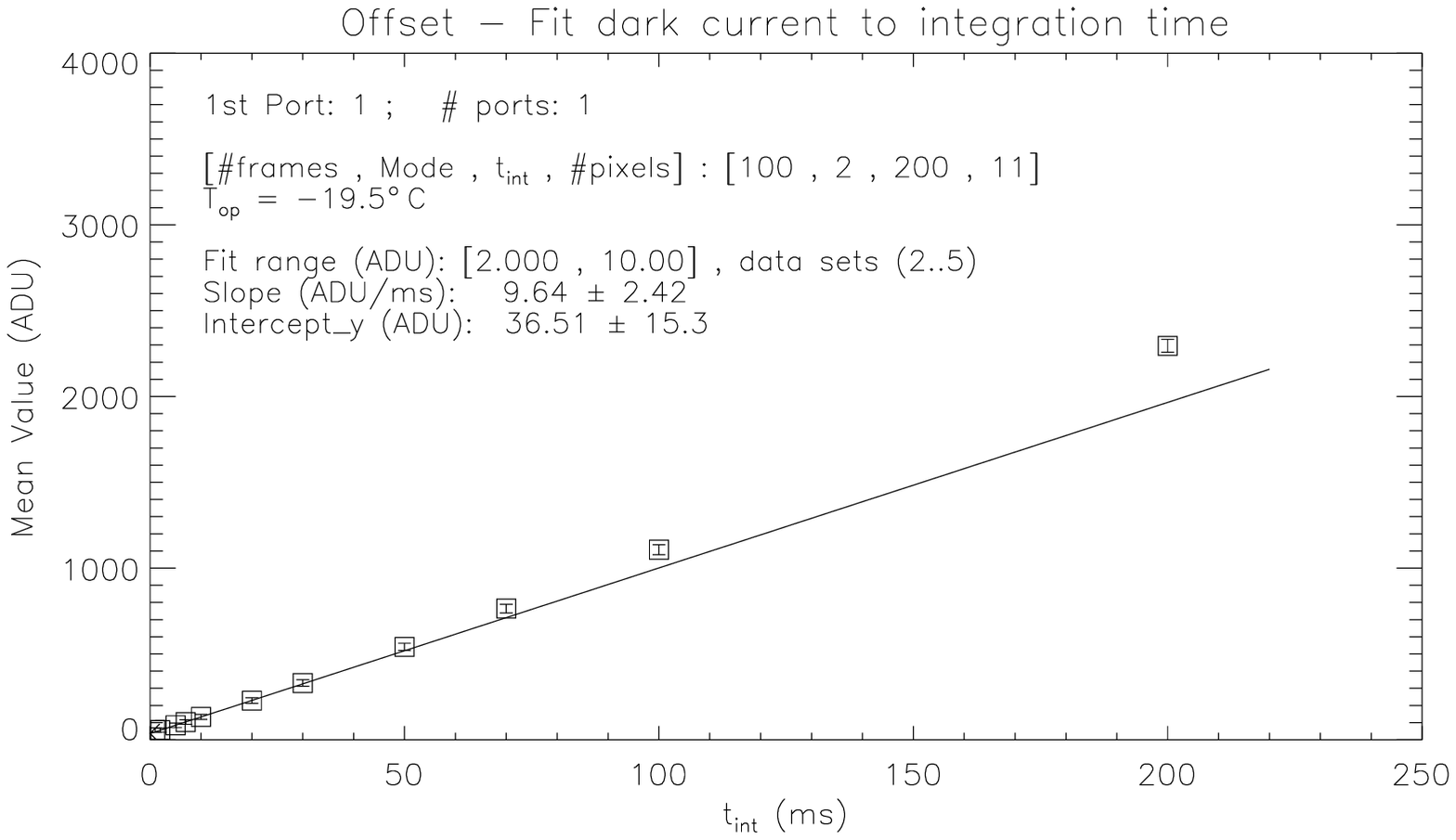}
        \includegraphics[width=0.49\textwidth]{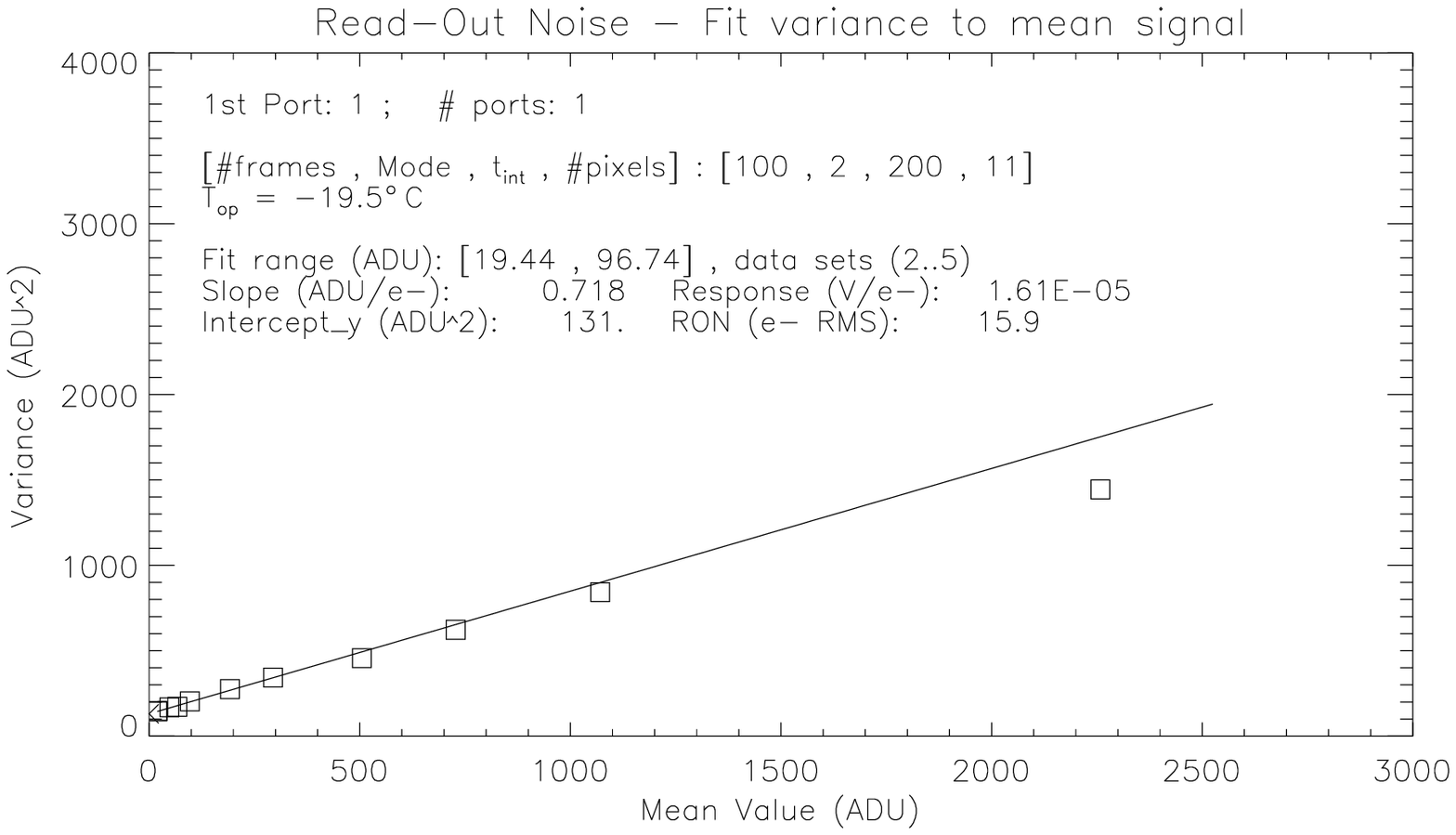}
    \caption{%
        \label{fig.darkcurr:noise}
        Retrieving bias (\emph{left}), RON, 
           and system gain (inverse slope, \emph{right}) in Mode~2
                (see Sect.~\ref{sect.ron-sysgain-m2}). %
        Plots are shown for all ports, $t_\mathrm{int}=1...200\,\mathrm{ms}$, 
        $T_\mathrm{op}=-19.5\grad$C; %
        fit range: data points 2..5, i.e., $t_\mathrm{int}=2..10\,\mathrm{ms}$.
        Data points average all pixels of one port at a given $t_\mathrm{int}$.
    }
    \end{figure}

This method was first applied to dark frames taken in Mode~1 
to verify validity and confirm the results found by the photon-transfer technique. 
Some results obtained for Mode~2, 
at the lowest temperature possible with the current setup ($T_\mathrm{op}=-19.5\grad$C), 
are shown in Figure~\ref{fig.darkcurr:noise}. 
They show that the system gain in Mode~2 is the same as in Mode~1, 
yet the read-out noise is roughly a factor 2 higher. 
In both modes the read-out noise corresponds to more than 2 bits of the AD converter, 
which is not optimal since this restrains its dynamic range. 
Also, the signal increases with integration time stronger than expected 
from the linear fit performed at lower signal levels. 
Non-linearity is roughly 9\% at $t_\mathrm{int}=100\unit{ms}$. 
The reason for this behavior is not yet clear and under investigation. 
In any case, this fact should have an only marginal effect on the operation. 

\begin{figure}
        \centering
        \includegraphics[width=0.49\textwidth]{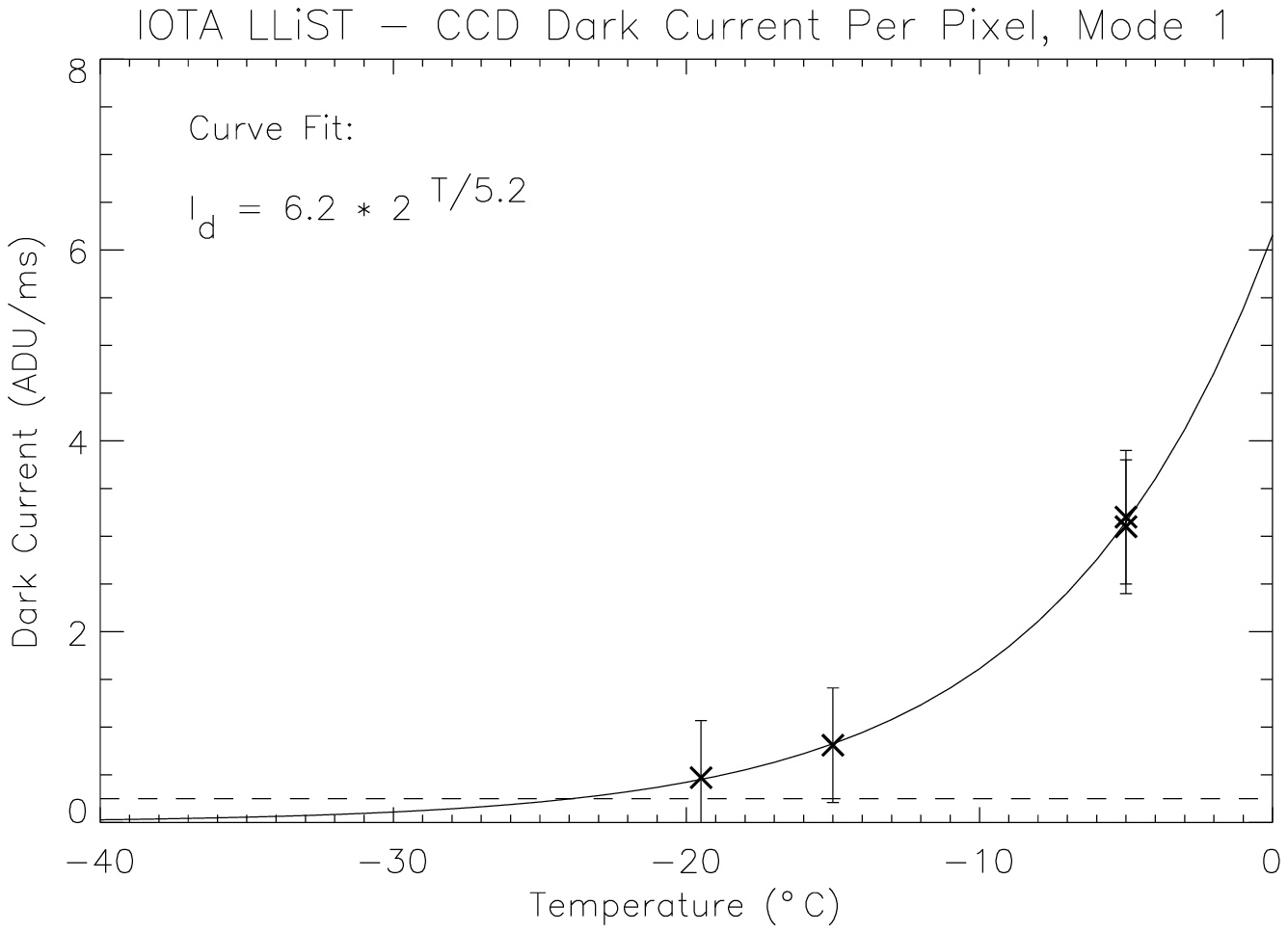} %
        \includegraphics[width=0.49\textwidth]{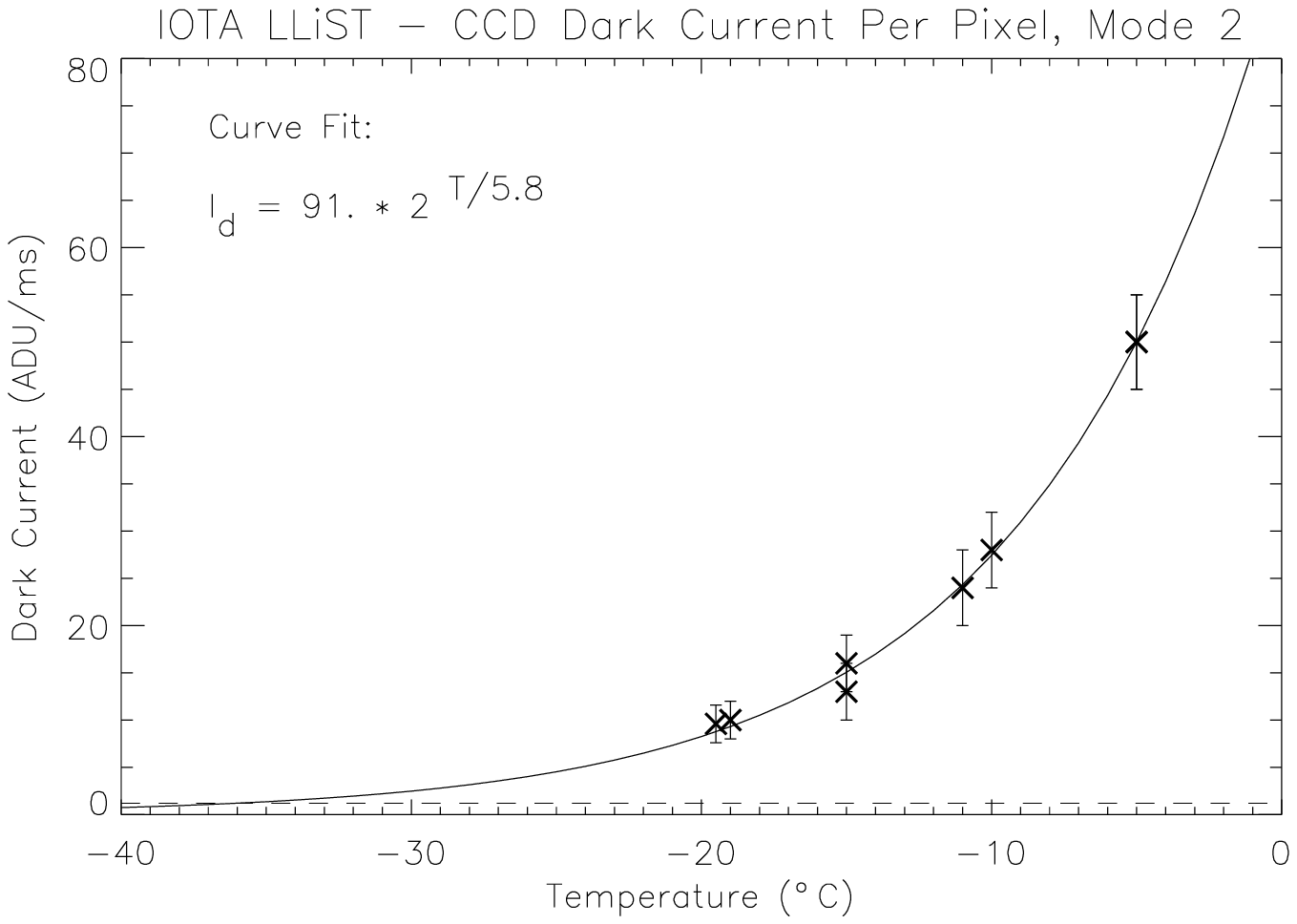} %
        \caption{%
        \label{fig.darkcurr:temper}
        Dark current of read-out modes M1 (left) and M2 (right). 
        The dashed lines near the bottom of the plots 
        indicate the dark current for which the photon noise (shot noise) 
        at an integration time $t_\mathrm{int}=100\unit{ms}$ 
        equals the read-out noise in the respective mode. 
        }
        \end{figure}

Figure~\ref{fig.darkcurr:temper} shows a compilation 
of the dark current obtained at different temperatures for Mode~1 and Mode~2 
and an exponential fit to them. 
The line at the bottom of the plots indicates the level 
where the photon noise originating from dark current 
at an integration time $t_\mathrm{int}=100\unit{ms}$ 
equals the read-out noise in the respective mode. 
It becomes apparent that it is necessary to cool down to $T_\mathrm{op}\approx -40\grad$C. 
This task will be addressed in the near future at IOTA
by attaching the fluid circuit of a glycol/water chiller 
to the heat sink of the CCD housing. 
The water chiller can reach temperatures of about $T\approx 0\grad$C. 
This allows the thermo-electric cooler to pump heat against a heat sink 
roughly $20\grad$C lower than currently room temperature 
and, therefore, reach the envisaged temperature.

\section{SUMMARY} 
\label{sect.summary}
The new CCD camera for the star tracker system at IOTA was installed earlier this year (2004) 
and has satisfactorily performed since. 
A summary of characteristics is given in Table~\ref{tab.summary}. 
        \begin{table}
        \caption{%
        \label{tab.summary}
        Summary of LLiST characteristics to date (June 2004). 
        For comparison, a selection of numbers is also given for the previous star tracker camera. 
        It operated in a mode very similar to LLiST Mode 2 ($4\times4$ binned). 
        }
        \begin{center}
        \begin{tabular}{lcc|c}
        \hline\hline
                  &  \multicolumn{2}{c|}{LLiST: CCD back illuminated}  &  Previous camera:  \\
                  &  Mode 1  &  Mode 2          &  CCD front illuminated  \\
                  &  (unbinned)  &  ($4\times4$ binned)  &  ($4\times4$ binned)  \\
        \hline
        Pixel size  &  21\unit{\mu m}  &  84\unit{\mu m}  &
                    \\
        Frame size  &  $64\times64\unit{pxs}$  &  $16\times16\unit{pxs}$  &
                  $8\times8\unit{pxs}$  \\
        Pixel scale on sky  &  0.72\unit{arcsec/px}  &  2.9\unit{arcsec/px}  &
                  $\sim4\unit{arcsec/px}$  \\
        Full FOV on sky (unvignetted)  &
                  \multicolumn{2}{c|}{$\sim 25\unit{arcsec}$}  &
                    \\
        Central spot of diffraction  &    &    &    \\
           \qquad limited star image at $\lambda_\mathrm{c}$  &
                  \multicolumn{2}{c|}{$23\unit{\mu m}$}  &
                    \\
        Maximum quantum efficiency  &
          \multicolumn{2}{c|}{90\% at $\lambda_\mathrm{c}=0.7\unit{\mu m}$}  &
                  40\% at 0.6\unit{\mu m}  \\
        Optical uniformity  &  \multicolumn{2}{c|}{$\pm5\%$}  &
                  $\sim\pm15\%$  \\
        Full well capacity  &  \multicolumn{2}{c|}{$\sim$175,000\,e-}  &
                    \\
        Fill factor  &  \multicolumn{2}{c|}{$\sim 100\%$}  &
                    \\
        Pixel clock-out rate  &  1.1\unit{MHz}  &  0.25\unit{MHz}  &    \\
        Maximum frame rate  &  796\unit{fps}  &  2220\unit{fps}  &
                  200\unit{fps}  \\
        Electronic gain $G_\mathrm{e}$  &    &    &    \\
           \qquad of pre-amplifier  &  \multicolumn{2}{c|}{6.8}  &
                    \\
        AD converter  &  \multicolumn{2}{c|}{2.5\unit{V} on 14 bit}  &
                    \\
            &  \multicolumn{2}{c|}{$K=(152.6\unit{\mu V/ADU})^{-1}$}  &
                    \\
        System gain $g_\mathrm{s}$  &
          \multicolumn{2}{c|}{$(1.5\pm0.2)$\,e-/ADU = $(0.67\pm0.1)^{-1}$\,e-/ADU}  &
                  $5.8$\,e-/ADU  \\
        Responsivity $R=(g_\mathrm{s}KG_\mathrm{e})^{-1}$  &    &  \\
          \qquad on CCD  &  \multicolumn{2}{c|}{$(15\pm1)\unit{\mu V/e-}$}  &
                    \\
        RON (rms)  &  $(7.5\pm1)$\,e-  &  $(16.5\pm1)$\,e-  &
                  $10.1$\,e-  \\
          &  $(5.0\pm1)$\,ADU  &  $(11.0\pm1)$\,ADU  &
                  $1.7$\,ADU  \\
        Linearity up to $t_\mathrm{int}=100\unit{ms}$  &
          \multicolumn{2}{c|}{$\pm5\%$}   &
                    \\
        Dark current at $-19.5\grad\mathrm{C}$  &
          $(0.47\pm0.6)\unit{ADU/ms}$  &  $(9.6\pm2)\unit{ADU/ms}$  &
                    \\
          \qquad (1 bright pixel)  &
          $(0.71\pm0.9)$\,e-/ms  &  $(14\pm3)$\,e-/ms  &
                    \\
          &  \multicolumn{2}{c|}{$26\unit{pA/cm^2}$}  &
                    \\
        Limiting magnitude  &
          \multicolumn{2}{c|}{?? [improvement by $\sim 2\unit{mag}$ expected]}  &
                  $I\sim 12\unit{mag}$  \\
        \hline
        \end{tabular}
        \end{center}
        \end{table}
The overall quantum efficiency of the CCD doubled, 
the frame rate of the camera increased by a factor 4 to 5 
up to the desired frame rate, %
the system gain improved by a factor 4, 
and read-out noise electrons were reduced by nearly a third 
operating in the unbinned Mode~1. 
Furthermore, first experience on the sky show an increase in sensitivity 
of roughly 1\unit{mag}, depending on spectral type of the star. 
Remaining issues are to be resolved: 
reducing high dark current to a non-disturbing level by additional cooling measures; 
lowering increased read-out noise in the binned Mode~2; 
balancing system gain and read-out noise 
for optimal use of the dynamic range of the AD converter. 
These tasks are expected to be accomplished in the summer down time 
before the start of the observing season in fall 2004. 
Other issues like the optimization of the IOTA system control software 
for the fast frame rates of LLiST will be addressed in the near future.

\acknowledgments     %

We thank all members of the Advanced Imaging Technology Group 
at MIT Lincoln Laboratory for their support 
under the leadership of Dr.\ Bernard B.\ Kosicki. 
\\[1ex]
The reported work was funded by the US National Science Foundation (NSF) 
under grant number AST-0138303 
to F.~P.~Schloerb (PI) at University of Massachusetts. 
\\[1ex]
This publication made use of 
NASA's Astrophysics Data System (ADS) Abstract Service 
(\url{http://adsabs.harvard.edu/}).

\bibliography{spie2004-pasch-arx}
\bibliographystyle{spiebib}   %

\end{document}

%% file: spie2004-copyright.tex
%
{
Copyright 2004 Society of Photo-Optical Instrumentation Engineers (SPIE). 
}
{
This paper was (will be) published in 
	{\em New Frontiers in Stellar Interferometry}, 
	W.~A. {Traub}, ed., 
	{\em SPIE Proceedings Series}, Vol.~{\bf 5491}, paper [5491-126].
and is made available as an electronic reprint (preprint) 
with permission of SPIE. 
One print or electronic copy may be made for personal use only. 
Systematic or multiple reproduction, 
distribution to multiple locations via electronic or other means, 
duplication of any material in this paper for a fee or for commercial purposes, 
or modification of the content of the paper are prohibited.
}